# A Profit-Sharing Mechanism for Coordinated Power-Transportation System

Tianyu Sima[a], Mingyu Yan[a], Jianfeng Wen[b*], Houbo Xiong[c*], Wensheng Luo[d], Mariusz Malinowski[e]

[a]*School of Electrical and Electronic Engineering, Huazhong University of Science and Technology, Wuhan 430074, China*
[b]*Department of Electrical Engineering and Electronics, University of Liverpool, L69 3GJ Liverpool, U.K.*
[c]*College of Electrical Engineering, Zhejiang University, Hangzhou 310027, China*
[d]*School of Electrical Engineering and Automation, Harbin Institute of Technology, Harbin 150001,China*
[e]*Institute of Control and Industrial Electronics, Warsaw University of Technology, 00-662 Warsaw, Poland*

A B S T R A C T

The transportation network operator (TNO) and the power distribution network operator (DNO) act non-cooperatively during the scheduling process. Under the TNO's management, the distribution of charging load may exacerbate the local supply-demand imbalance in the power distribution network (PDN), which negatively impacts the secure and economic operation of the PDN. This paper proposes a profit-sharing mechanism based on the principle of incentive compatibility for coordinating the transportation network (TN) and the PDN to minimize the total operation cost of the PDN. In this mechanism, the scheduling process of the power-transportation system is divided into two stages. At the pre-scheduling stage, the TNO allocates traffic flow and charging load without considering the operation of the PDN, after which the DNO schedules and obtains the original cost. At the re-scheduling stage, the DNO shares part of the saved dispatch cost to motivate the TNO to reallocate the EVs' charging, which is more beneficial to the operation of the PDN. This two-stage process is then simulated by two single-level models and a bilevel model. Finally, the optimal sharing ratio is identified, at which the total scheduling cost of the DNO can decrease to the lowest point when gaming with the TNO. The efficiency of the proposed mechanism is simulated via a coupled network with 12 traffic nodes and 18 electric buses. Numerical results demonstrate that the DNO can achieve the minimum total cost. Simultaneously, the TNO can also benefit from the proposed profit-sharing mechanism.

## 1. Introduction

With the worsening global environment and the scarcity of fossil fuels, there is a growing desire to replace traditional fuel-powered vehicles with electric vehicles (EVs), which are environmentally friendly alternatives. The International Energy Agency (IEA) released the Global EV Outlook 2024, which highlights that the global EV penetration rate is projected to reach two-thirds by 2035 [1]. It is concerning that the EV charging load with random spatiotemporal distribution characteristics poses challenges to the operation of the power distribution network (PDN). At the temporal level, EVs charge during peak electricity demand periods, further exacerbating the peak-valley difference, intensifying the peak shaving pressure on the PDN. At the spatial level, a large number of EVs charge at a specific electric vehicle charging station (EVCS) may intensify the electric supply-demand imbalance in the PDN [2], potentially causing issues such as overloads of distribution lines, voltage drop [3]-[4], increased losses in the PDN [5]-[6], etc. Traditionally, the PDN can be enhanced to overcome these challenges from both planning and scheduling perspectives, such as expanding electric line/ transformer capacity, deploying energy storage systems/ charging stations [7]-[9], and dispatching distributed emergency generators, etc. However, these cost-ineffective methods cannot solve the problem fundamentally. Strengthening the management of the EVs to optimize the distribution of charging load itself seems to be a more effective way to ensure the efficient and stable operation of the PDN.

Currently, the transportation network operator (TNO) utilizes two major types of traffic simulation models to manage EVs. One is the system optimization (SO)-based model, the other is the user equilibrium (UE)-based model [10]-[11]. For the former, all vehicles in the transportation network (TN) are assumed to fully adhere to the instructions of the TNO for minimizing the TN's operating cost [12]. However, this model is overly idealized. While it ensures the system optimization, the traveling time for some users is increased. Actually, most self-interested drivers are unlikely to sacrifice for the overall benefit of the TN and prefer to choose a route with the lowest travel cost. The traveling characteristics of users can be well described by the UE model [13]-[15]. On this basis, the TNO can formulate suitable policies to guide EVs' route choice and charging strategy to improve the efficiency of TN. The existing research shows that the implementation of road congestion tolls and adjustments to charging service fees at EVCSs are two possible measures to guide EV users, which have been proven to be effective in some countries, such as the UK and Singapore.

However, due to the non-cooperative relationship between the TN and the PDN, the TNO usually regulates the traffic flow and charging load without considering the possible negative impact on the operation of the PDN. As the penetration rate of EVs gradually increases, researchers in the field of power system have garnered increasing attention to this problem [16]-[22]. In recent years, some methods have been proposed to coordinate the operation of the power and transportation systems, which improves the economy and security of the PDN. In [23], a holistic framework is proposed to enhance the operation of the PDN and the TN via EV charging services. It's described by a bilevel model and solved by a deep reinforcement learning-based algorithm. A multi-period optimal traffic and power flow model is formulated in [24], where the distribution of traffic flow is represented by a semi-dynamic traffic assignment model. Then, a non-profit system operator is assumed to use this model to set road congestion tolls and dispatch local generators to minimize the total operation cost of

**Nomenclature**

*Abbreviations*

|  |  |
|---|---|
| TN | Transportation network |
| PDN | Power distribution network |
| TNO | Transportation network operator |
| DNO | Power distribution network operator |
| EVCS | Electric vehicle charging station |
| LG | Local emergency generator |

*Indices, Subscripts, and Sets*

|  |  |
|---|---|
| $a$, $m$, $rs$ | Index of roads, electric vehicle charging stations, and O-D pairs |
| $i$ | Index of local emergency generators |
| road, EVCS | Subscript for road and electric vehicle charging station |
| GV, EV | Subscript for gasoline and electric vehicle |
| LG, sub | Subscript for local generators and substation |
| $\rho_{rs}$ | Set of routes for O-D pair $rs$ |
| $\mathcal{R}$ , $\mathcal{E}$ | Set of roads and charging stations |

*Variables*

|  |  |
|---|---|
| $\mathbf{x}$ , $\mathbf{y}$ | Column vector of traffic flow on roads and at EVCSs |
| $\mathbf{x}^{\mathrm{GV/EV}}$ | Column vector of GV / EV traffic flow on roads |
| $\mathbf{f}^{\mathrm{GV/EV}}$ | Column vector of GV / EV traffic flow on routes |
| $\mathbf{t}^{\mathrm{road/EVCS}}$ | Column vector of travel time on roads / charging time at EVCSs |
| $\mathbf{C}^{\mathrm{road/EVCS}}$ | Column vector of overall cost on roads / at EVCSs |
| $\mathbf{C}^{\mathrm{GV/EV}}$ | Column vector of GV/EV travel cost on routes |
| $\mathbf{T}^{\mathrm{R/E}}$ | Column vector of congestion toll on roads / additional entry fee at EVCSs |
| $\mathbf{u}^{\mathrm{GV/EV}}$ | Column vector of GV / EV optimal travel cost for O-D pairs |
| $\mathbf{P}^{\mathrm{EVCS}}$ | Column vector of charging load at EVCSs |
| $\mathbf{p}^{\mathrm{LG}}$ / $P^{\mathrm{sub}}$ | Column vector of the local generators' / substation's power generation |
| $\mathbf{F}^{\mathrm{LG}}$ / $F^{\mathrm{sub}}$ | Column vector of the local generators' / substation's power generation cost |
| $\mathbf{P}^{\mathrm{l}}$ | Column vector of active power of electric lines |
| $\boldsymbol{\theta}$ | Column vector of the voltage angle of buses |
| $\Gamma$ | Total user's travel cost of TN |
| $\eta$ | Power dispatch cost of DNO |
| $\alpha$ | Sharing ratio from the DNO to the TNO |

*Parameters*

|  |  |
|---|---|
| $\mathbf{q}^{\mathrm{GV/EV}}$ | Column vector of GV / EV traffic demand for O-D pairs |
| $\boldsymbol{\Lambda}^{\mathrm{GV/EV}}$ | Incidence matrix between O-D pairs and routes for GV / EV |
| $\boldsymbol{\delta}^{\mathrm{GV/EV}}$ | Incidence matrix between roads and routes for GV / EV |
| $\boldsymbol{\gamma}$ | Incidence matrix between EVCSs and routes of EV |
| $\mathbf{c}^{\mathrm{road/EVCS}}$ | Column vector of roads' / EVCSs' capacity |
| $t_a^{\mathrm{road},0}$ / $t_m^{\mathrm{EVCS},0}$ / $J$ | Empirical coefficients in road traveling time and EVCS charging time estimation |
| $\omega$ | Monetary value of time |
| $\zeta$ | Charging price at EVCSs |
| $E_B$ | The charging demand of each EV |
| $\mathbf{G}$ / $\mathbf{D}$ / $\mathbf{L}$ | Incidence matrix between nodes and generation units / EVCSs / electric lines |
| $\mathbf{p}^{\mathrm{td}}$ | Traditional electric demands at each bus |
| $\mathbf{X}$ | Column vector of the reactance of electric lines |

the power-transportation system. As an extension of this work, the hydrogen energy storage (HES) system is considered, which brings more flexibility for the operation of the power-transportation system [25]. Although these frameworks contribute to the overall efficiency of the power-transportation system, the profit allocation process is not considered. As a result, the distribution network operator (DNO) may benefit unilaterally while the TNO has no motivation to regulate the EV charging load for the efficient operation of the PDN. Therefore, it is necessary to study a suitable profit allocation scheme to evoke the adjustment potential of the traffic side.

The principle of incentive compatibility has been applied in various fields such as financial regulation, corporate management, and policy-making. It allows participants to maximize their interests in a way that aligns with the strategies desired by the mechanism designer [26]. Inspired by this theory, this paper proposes a profit-sharing mechanism for the power-transportation system. It motivates the TNO voluntarily to leverage the spatial flexibility of EV charging load to promote the economic operation of the PDN. In general, the contributions of this paper are summarized as follows:

(1) A profit-sharing mechanism is proposed to promote the coordination between the PDN and the TN. Within this mechanism, the DNO shares part of the saved dispatching cost with the TNO. Driven by the benefit, the TNO reallocates the distribution of EV charging load to alleviate the local supply-demand imbalance in the PDN to promote a more efficient operation of the PDN. Under this mechanism, the benefit of the DNO can be fully secured while the TNO can also obtain a considerable benefit.

(2) The optimal traffic flow allocation model and power economic dispatch model are established, respectively. The two single-level models are utilized to simulate the decision-making of the TNO and the DNO at the pre-scheduling stage. A bilevel mathematical model is formulated to reflect the strategy adjustment of the TNO and the DNO when the profit-sharing is introduced, i.e., the situation at the re-scheduling stage. In this article, the lower level is transferred to the constraints in the upper level via the KKT condition to solve the bilevel model. After the two-stage simulation, the expected operation situations of the TN and the PDN are obtained.

(3) The technique of linearization is employed to convert the above-mentioned nonlinear models into linear programming (LP) models. With such a linearization technique, the accuracy of the models can be guaranteed, and the solving complexity of the models is decreased, enabling stable optimization and solution. Finally, according to the simulation results based on these models, the optimal sharing ratio is determined through sensitivity analysis.

The rest of this article is organized as follows. Section 2 presents the framework for profit-sharing between power and transportation systems. Section 3 provides the scheduling model of the power-transportation system at the pre-scheduling stage. Section 4 provides the scheduling model of the power-transportation system at re-scheduling stage and the method to determine the best profit-sharing ratio. Section 5 illustrates the linearization of nonlinear constraints. Section 6 provides simulation results based on a coupled power-transportation system. Section 7 draws the conclusion.

## 2. Framework of profit sharing for the power-transportation system

This part illustrates the overall profit-sharing mechanism based on the principle of incentive compatibility. The specific scheduling process for the power and transportation systems under this mechanism is illustrated in Fig. 1. It is divided into two stages as pre-scheduling and re-scheduling.

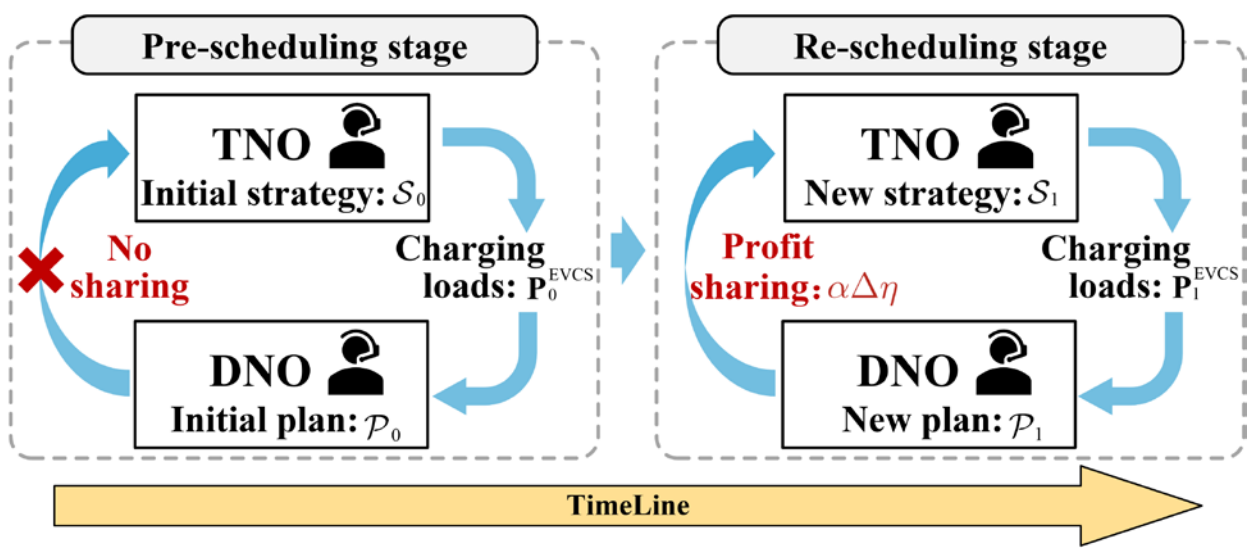

**Fig. 1.** Illustration of the scheduling process of the power-transportation system under the profit-sharing mechanism

At the pre-scheduling stage, the TNO allocates the traffic flow and charging load without considering the operation of the PDN. Its objective is only to minimize the total travel cost of users and develop the initial regulation strategy as $\mathcal{S}_0$. In this context, the predicted charging load is $\mathbf{P}_0^{\text{EVCS}}$. Based on the forecast of traditional electric demand $\mathbf{p}^{\text{td}}$ and charging load $\mathbf{P}_0^{\text{EVCS}}$, the DNO creates the initial economic dispatch plan $\mathcal{P}_0$ and calculates the benchmark operation cost $\eta_0$. At this stage, there may be a serious mismatch between the electric supply and demand at some buses. The DNO has to dispatch the local generators with high marginal costs to guarantee the power balance. Subsequently, the operation cost of the PDN is high.

At the re-scheduling stage, if the TNO reallocates the charging load to help the DNO decrease the dispatch cost, the DNO will share a part of the saved money with the TNO. However, the reallocation must result in an increased total travel cost in the TN. After weighing the shared money from the DNO and the loss of users, the TNO makes the decision as $\mathcal{S}_1$ that minimizes the total operation cost of the TN. Then, the DNO re-schedules as $\mathcal{P}_1$ and the new dispatch cost of the DNO is $\eta_1$.

The operation of the PDN and the TN is shown in Fig. 2. Under the mechanism, the mismatch between electric supply and demand is alleviated in PDN by redistributing the charging load. Therefore, the DNO can reduce the utilization of local emergency generators with high marginal costs after rescheduling. The dispatch cost of the DNO is decreased from $\eta_0$ to $\eta_1$. The corresponding shared amount of money for the TNO is $\alpha(\eta_0-\eta_1)$, where $\alpha$ is the sharing ratio. Finally, the security and economy of the PDN are enhanced, while the benefit of the TN is safeguarded.

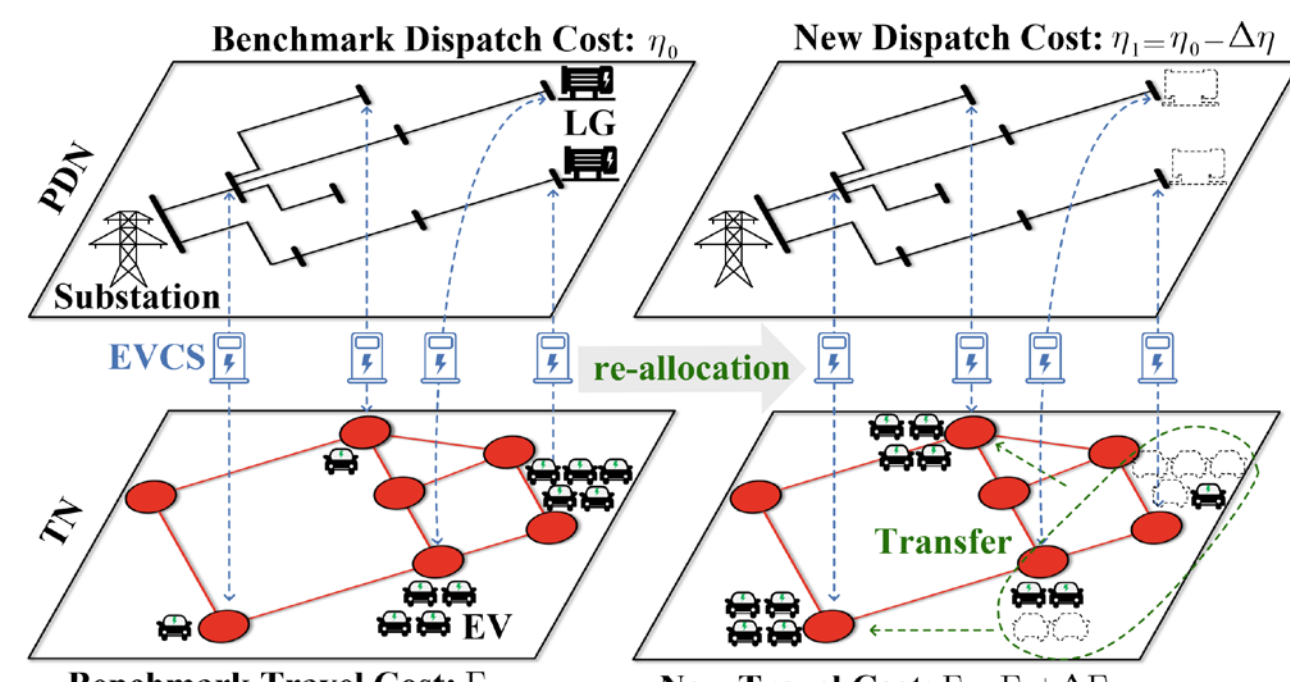

**Fig. 2.** Power generation and dispatch cost in PDN & charging load distribution and travel cost in TN at the two stages

## 3. Scheduling model of power-transportation system in the pre-scheduling stage

### *3.1. Optimal Traffic Flow Allocation Model*

The TN can be abstractly represented by a connected directed graph $\mathcal{G}=\{\mathcal{N},\mathcal{A}\}$, where $\mathcal{N}$ and $\mathcal{A}$ represent node and road set, respectively [27]. Each vehicle on TN drives from the origin node to the destination node (O-D pair). For O-D pair *rs*, it's connected by multiple routes, which form the route set $\rho_{rs}$. Based on the network constraints and the Wardrop UE principle, O-D pair traffic demand $q_{rs}$ is assigned to routes in $\rho_{rs}$.

In this article, the TNO manages the distribution of traffic flow and charging load by imposing additional road congestion tolls and charging station entry fees. At the pre-scheduling stage, the objective of the TNO is just to minimize the total travel costs of all vehicles on the TN. The corresponding mathematical form is formulated as:

$$\min_{\mathbf{T}^{\text{R}},\mathbf{T}^{\text{E}}} \Gamma \tag{1a}$$

$$\begin{aligned}\Gamma &= (\mathbf{C}^{\text{road}})^{\text{T}}\mathbf{x}+(\mathbf{C}^{\text{EVCS}})^{\text{T}}\mathbf{y} \\ &= (\mathbf{u}^{\text{GV}})^{\text{T}}\mathbf{q}^{\text{GV}}+(\mathbf{u}^{\text{EV}})^{\text{T}}\mathbf{q}^{\text{EV}}\end{aligned} \tag{1b}$$

The operation constraints of TN are formulated as (2)-(18). Constraints (2)-(3) ensure that O-D pair traffic demands $\mathbf{q}^{\text{GV}}$ and $\mathbf{q}^{\text{EV}}$ are balanced by route traffic flow $\mathbf{f}^{\text{GV}}$ and $\mathbf{f}^{\text{EV}}$, respectively. Constraints (4)-(5) describes that GV/EV road traffic flow is the sum of GV/EV route traffic flow that passes this road. Constraint (6) represents that the total road traffic flow consists of these two types of traffic. Constraint (7) states that EVCS traffic flow is the sum of EV route traffic flow that charges at the respective EVCS. Constraint (8) sets the traffic capacity of roads and EVCSs. Constraint (9) establishes the proportional relationship between the charging load and EVCS traffic flow by disregarding the heterogeneity of EVs. Constraint (10) utilizes Bureau of Public Roads function to estimate the road traveling time [28]. Based on queuing theory [29], constraint (11) models the time spent at the EVCS. Constraint (12) corresponds to the cost on a road, including time cost and congestion toll. Constraint (13) corresponds to the cost for EVs at an EVCS, including time cost, charging fee, and additional station entry fee. Constraint (14) is the non-negative condition of congestion toll and additional station entry fee. Constraints (15)-(16) calculate the travel costs for GVs and EVs on the routes. Constraints (17)-(18) describe the equilibrium state of the TN: no driver can reduce its travel cost further by changing the route selection unilaterally.

$$\mathbf{q}^{\text{GV}}=\mathbf{\Lambda}^{\text{GV}}\mathbf{f}^{\text{GV}} \tag{2}$$

$$\mathbf{q}^{\text{EV}}=\mathbf{\Lambda}^{\text{EV}}\mathbf{f}^{\text{EV}} \tag{3}$$

$$\mathbf{x}^{\text{GV}}=\mathbf{\delta}^{\text{GV}}\mathbf{f}^{\text{GV}} \tag{4}$$

$$\mathbf{x}^{\mathrm{EV}} = \boldsymbol{\delta}^{\mathrm{EV}} \mathbf{f}^{\mathrm{EV}} \tag{5}$$

$$\mathbf{x} = \mathbf{x}^{\mathrm{GV}} + \mathbf{x}^{\mathrm{EV}} \tag{6}$$

$$\mathbf{y} = \boldsymbol{\gamma} \mathbf{f}^{\mathrm{EV}} \tag{7}$$

$$\mathbf{x} \leq \mathbf{c}^{\mathrm{road}}, \ \ \mathbf{y} < \mathbf{c}^{\mathrm{EVCS}} \tag{8}$$

$$\mathbf{P}^{\mathrm{EVCS}} = \mathbf{y} E_B \tag{9}$$

$$\mathbf{t}^{\mathrm{road}}(\mathbf{x}) = \left[ t_a^{\mathrm{road},0} \left( 1 + 0.15 \left( \frac{x_a}{c_a^{\mathrm{road}}} \right)^4 \right) \right]_{a \in \mathcal{R}}^{\mathrm{T}} \tag{10}$$

$$\mathbf{t}^{\mathrm{EVCS}}(\mathbf{y}) = \left[ t_m^{\mathrm{EVCS},0} \left( 1 + J \left( \frac{y_m}{c_m^{\mathrm{EVCS}} - y_m} \right) \right) \right]_{m \in \mathcal{E}}^{\mathrm{T}} \tag{11}$$

$$\mathbf{C}^{\mathrm{road}} = \varpi \mathbf{t}^{\mathrm{road}} + \mathbf{T}^{\mathrm{R}} \tag{12}$$

$$\mathbf{C}^{\mathrm{EVCS}} = (\varpi \mathbf{t}^{\mathrm{EVCS}} + \mathbf{1} \zeta E_B) + \mathbf{T}^{\mathrm{E}} \tag{13}$$

$$\mathbf{T}^{\mathrm{R}} \geq \mathbf{0}, \ \mathbf{T}^{\mathrm{E}} \geq \mathbf{0} \tag{14}$$

$$\mathbf{C}^{\mathrm{GV}} = (\boldsymbol{\delta}^{\mathrm{GV}})^{\mathrm{T}} \mathbf{C}^{\mathrm{road}} \tag{15}$$

$$\mathbf{C}^{\mathrm{EV}} = (\boldsymbol{\delta}^{\mathrm{EV}})^{\mathrm{T}} \mathbf{C}^{\mathrm{road}} + \boldsymbol{\gamma}^{\mathrm{T}} \mathbf{C}^{\mathrm{EVCS}} \tag{16}$$

$$\mathbf{0} \leq \mathbf{f}^{\mathrm{GV}} \perp \left[ \mathbf{C}^{\mathrm{GV}} - (\boldsymbol{\Lambda}^{\mathrm{GV}})^{\mathrm{T}} \mathbf{u}^{\mathrm{GV}} \right] \geq \mathbf{0} \tag{17}$$

$$\mathbf{0} \leq \mathbf{f}^{\mathrm{EV}} \perp \left[ \mathbf{C}^{\mathrm{EV}} - (\boldsymbol{\Lambda}^{\mathrm{EV}})^{\mathrm{T}} \mathbf{u}^{\mathrm{EV}} \right] \geq \mathbf{0} \tag{18}$$

### *3.2. Power Economic Dispatch Model*

The DNO can purchase electricity from the wholesale market and dispatch the local energy resources to balance the electric load. At the pre-scheduling stage, its objective is to minimize the total dispatch costs, and the corresponding mathematical form is formulated as:

$$\min_{\mathbf{P}^{\mathrm{G}}} \ \ \eta \tag{19}$$

$$\eta = \sum_i \left[ a_i (p_i^{\mathrm{LG}})^2 + b_i p_i^{\mathrm{LG}} + c_i \right] + \kappa P^{\mathrm{sub}} = \mathbf{1}^{\mathrm{T}} \mathbf{F}^{\mathrm{G}} \quad : \lambda^1 \tag{20}$$

$$\mathbf{F}^{\mathrm{G}} \geq \mathbf{k}_j \odot \mathbf{P}^{\mathrm{G}} + \mathbf{b}_j \quad : \boldsymbol{\mu}_j^0, \quad \forall j \tag{21}$$

where the overall operation cost of the PDN is given by constraint (20). The first term is the fuel cost of local generators. The second term is the cost of purchasing electricity from the wholesale market, where $\kappa$ is the electricity price. In constraint (21), the piecewise-linear function is used to approximate the nonlinear operation cost [30]. For notational conciseness, we defines $\mathbf{F}^{\mathrm{G}} = [\mathbf{F}^{\mathrm{LG}}, F^{\mathrm{sub}}]$, $\mathbf{P}^{\mathrm{G}} = [\mathbf{p}^{\mathrm{LG}}, P^{\mathrm{sub}}]$, $\mathbf{k}_j = [k_{1,j}, \ldots, k_{NG,j}, k_{NG+1,j}]$, $\mathbf{b}_j = [b_{1,j}, \ldots, b_{NG,j}, b_{NG+1,j}]$, where *NG* is the number of the local generators and *j* represents the interval number of the piecewise function. $\lambda^1$ and $\boldsymbol{\mu}_j^0$ are the dual variables of constrains (20) and (21), respectively.

Constraints (22)-(26) are the security operation constraints of PDN. Constraint (22) ensures the power balance at each bus. In constraint (23), the active power of the electric line is formulated by the power flow model mentioned in [31] and [32]. Constraint (24) gives the bound of the generator output. Constraint (25) provides capacity for lines. Constraint (26) limits the variation of the bus angle. The dual variables of constraint (22)-(23) are $\boldsymbol{\lambda}^2$, $\boldsymbol{\lambda}^3$, $\underline{\boldsymbol{\mu}}^1, \overline{\boldsymbol{\mu}}^1$, $\underline{\boldsymbol{\mu}}^2, \overline{\boldsymbol{\mu}}^2$ and $\underline{\boldsymbol{\mu}}^3, \overline{\boldsymbol{\mu}}^3$, respectively.

$$\mathbf{G} \mathbf{P}^{\mathrm{G}} - (\mathbf{p}^{\mathrm{td}} + \mathbf{D} \mathbf{P}^{\mathrm{EVCS}}) = \mathbf{L} \mathbf{P}^{\mathrm{l}} \quad : \boldsymbol{\lambda}^2 \tag{22}$$

$$\mathbf{P}^{\mathrm{l}} \odot \mathbf{X} = \mathbf{L}^{\mathrm{T}} \boldsymbol{\theta} \quad : \boldsymbol{\lambda}^3 \tag{23}$$

$$\mathbf{P}_{\min}^{\mathrm{G}} \leq \mathbf{P}^{\mathrm{G}} \leq \mathbf{P}_{\max}^{\mathrm{G}} \quad : \underline{\boldsymbol{\mu}}^1, \overline{\boldsymbol{\mu}}^1 \tag{24}$$

$$-\mathbf{P}_{\max}^{\mathrm{l}} \leq \mathbf{P}^{\mathrm{l}} \leq \mathbf{P}_{\max}^{\mathrm{l}} \quad : \underline{\boldsymbol{\mu}}^2, \overline{\boldsymbol{\mu}}^2 \tag{25}$$

$$\boldsymbol{\theta}_{\min} \leq \boldsymbol{\theta} \leq \boldsymbol{\theta}_{\max} \quad : \underline{\boldsymbol{\mu}}^3, \overline{\boldsymbol{\mu}}^3 \tag{26}$$

## 4. Scheduling model of power-transportation system at the re-scheduling stage

In this section, the classic Stackelberg model is utilized to simulate the game between the TNO and the DNO at re-scheduling stage. It's a bilevel optimization problem described as (27)-(30). At the upper level, the distribution of traffic flow and charging load is optimized by the TNO. The DNO develops the economic dispatching at the lower level.

### *4.1. Upper Level: Optimal Traffic Flow Allocation Model Considering Profit Sharing*

$$\min_{\mathbf{T}^{\mathrm{R}}, \mathbf{T}^{\mathrm{E}}} \ \ \Gamma - \alpha(\eta_0 - \eta) \tag{27}$$

$$\text{Constraints (2)-(18)} \tag{28}$$

### 4.2. *Lower* Level: Power Economic Dispatch Model Considering Profit Sharing

$$\min_{\mathbf{P}^{\mathrm{G}}} \ \ \eta + \alpha(\eta_0 - \eta) \tag{29}$$

$$\text{Constraints (20)-(26)} \tag{30}$$

Compared to the models mentioned in Section 3, the difference is only reflected in the objective function. Constraint (27) is the new objective of TNO. Under the sharing mechanism, the TNO needs to consider the profit $\alpha(\eta_0 - \eta)$ obtained from the DNO when making decisions. The operation constraints of the TN remain unchanged, as formulated in constraint (28). The DNO's objective (29) is modified to minimize the power generation cost plus the profit sharing. The operation constraints of the PDN remain the same, expressed in constraint (30).

### *4.3. Determining the optimal sharing ratio*

The optimal sharing ratio is defined as the point where the PDN total operation cost $\Psi(\alpha) = \eta + \alpha(\eta_0 - \eta)$ is the least:

$$\alpha^* = \arg\min \Psi(\alpha) \tag{31}$$

The solution result of bilevel optimization problem (27)-(30) can serve as the basis for selecting the sharing ratio. However, the multiplication of continuous variables $\alpha$ and $\eta$ leads to the nonlinear constraints (27) and (29), making it difficult to solve this problem. Here we gradually increase $\alpha$ from 0 to 1 and fix it during each solution. Since $\eta_0$ and $\alpha$ are constants during each solution, the objective function (29) $\min \ \ (1-\alpha)\eta + \alpha\eta_0$ can be simplified to $\min \ \ \eta$, which is the same as (19). As a result, the lower-level problem can be equivalently described by the model established in section 3.2. This linear problem can be converted into equilibrium constraints in the upper-level problem using Karush-Kuhn-Tucker (KKT) conditions. Ultimately, the bilevel optimization is transformed into a single-level optimization as follows:

$$\min_{\mathbf{T}^{\mathrm{R}}, \mathbf{T}^{\mathrm{E}}} \ \ \Gamma - \alpha(\eta_0 - \eta) \tag{32}$$

$$\text{Constraints (2)-(18)} \tag{33}$$

$$1 + \lambda^1 = 0 \tag{34}$$

$$-\mathbf{1} \lambda^1 - \sum_j \boldsymbol{\mu}_j^0 = \mathbf{0} \tag{35}$$

$$\sum_j \boldsymbol{\mu}_j^0 \odot \mathbf{k}_j + \mathbf{G}^{\mathrm{T}} \boldsymbol{\lambda}^2 - \underline{\boldsymbol{\mu}}^1 + \overline{\boldsymbol{\mu}}^1 = \mathbf{0} \tag{36}$$

$$-\mathbf{L}^{\mathrm{T}} \boldsymbol{\lambda}^2 + \boldsymbol{\lambda}^3 \odot \mathbf{X} - \underline{\boldsymbol{\mu}}^2 + \overline{\boldsymbol{\mu}}^2 = \mathbf{0} \tag{37}$$

$$-\mathbf{L}\boldsymbol{\lambda}^{3}-\underline{\boldsymbol{\mu}}^{3}+\overline{\boldsymbol{\mu}}^{3}=\mathbf{0} \tag{38}$$

$$\mathbf{0}\leq\boldsymbol{\mu}_{j}^{0}\perp(-\mathbf{F}^{\mathrm{G}}+\mathbf{k}_{j}\odot\mathbf{P}^{\mathrm{G}}+\mathbf{b}_{j})\leq\mathbf{0} \tag{39}$$

$$\mathbf{0}\leq\underline{\boldsymbol{\mu}}^{1}\perp(-\mathbf{P}^{\mathrm{G}}+\mathbf{P}_{\min}^{\mathrm{G}})\leq\mathbf{0} \tag{40}$$

$$\mathbf{0}\leq\overline{\boldsymbol{\mu}}^{1}\perp(\mathbf{P}^{\mathrm{G}}-\mathbf{P}_{\max}^{\mathrm{G}})\leq\mathbf{0} \tag{41}$$

$$\mathbf{0}\leq\underline{\boldsymbol{\mu}}^{2}\perp(-\boldsymbol{P}^{\mathrm{l}}-\mathbf{P}_{\max}^{\mathrm{l}})\leq\mathbf{0} \tag{42}$$

$$\mathbf{0}\leq\overline{\boldsymbol{\mu}}^{2}\perp(\mathbf{P}^{\mathrm{l}}-\mathbf{P}_{\max}^{\mathrm{l}})\leq\mathbf{0} \tag{43}$$

$$\mathbf{0}\leq\underline{\boldsymbol{\mu}}^{3}\perp(-\boldsymbol{\theta}+\boldsymbol{\theta}_{\min})\leq\mathbf{0} \tag{44}$$

$$\mathbf{0}\leq\overline{\boldsymbol{\mu}}^{3}\perp(\boldsymbol{\theta}-\boldsymbol{\theta}_{\max})\leq\mathbf{0} \tag{45}$$

$$\eta-\mathbf{1}^{\mathrm{T}}\mathbf{F}^{\mathrm{G}}=0 \tag{46}$$

$$\mathbf{G}\mathbf{P}^{\mathrm{G}}-(\mathbf{p}^{\mathrm{td}}+\mathbf{D}\mathbf{P}^{\mathrm{EVCS}})=\mathbf{L}\mathbf{P}^{\mathrm{l}} \tag{47}$$

$$\mathbf{P}^{\mathrm{l}}\odot\mathbf{X}=\mathbf{L}^{\mathrm{T}}\boldsymbol{\theta} \tag{48}$$

Constraint (33) describes the equilibrium state of the TN. Constraints (34)-(38) represent the stationarity conditions, which are the partial derivatives of variables $\eta$ , $\mathbf{F}^{\mathrm{G}}$ , $\mathbf{P}^{\mathrm{G}}$ , $\mathbf{P}^{\mathrm{l}}$ , $\boldsymbol{\theta}$ . Constraints (39)-(48) are the equilibrium constraints, which represent the complementary slackness, primal feasibility, and dual feasibility.

## 5. Linearization of constraints

### *5.1. Linearization of Constraint (10)*

Constraint (10) represents a nonlinear time function related to road traffic flow. The piecewise linear function can be used to approximate it as shown in Fig. 3. On time function of road a, $N$+1 points at equal intervals ( $\Delta x_a^{\max}$ ) is selected along the horizontal axis. Then, each pair of adjacent points is connected to form $N$ line segments. The slope of segment n is denoted as $A_{a,j}$ . $\Delta x_{a,j}$ indicates the width that the horizontal axis interval $j$ is filled. For notational conciseness, this paper defines $\Delta\mathbf{x}^{\max}=[\Delta x_a^{\max}]_{a\in\mathcal{R}}$ , $\mathbf{A}_j=[A_{a,j}]_{a\in\mathcal{R}}$ , $\Delta\mathbf{x}_j=[\Delta x_{a,j}]_{a\in\mathcal{R}}$ , $\mathbf{Z}_j=[Z_{a,j}]_{a\in\mathcal{R}}$ .

$$\mathbf{x}=\sum_{j=1}^{N}\Delta\mathbf{x}_j \tag{49}$$

$$\mathbf{t}^{\mathrm{road}}=\mathbf{t}^{\mathrm{road},0}+\sum_{j=1}^{N}\mathbf{A}_j\odot\Delta\mathbf{x}_j \tag{50}$$

$$\Delta\mathbf{x}^{\max}=\mathbf{c}^{\mathrm{road}}/N \tag{51}$$

$$\mathbf{0}\leq\Delta\mathbf{x}_j\leq\Delta\mathbf{x}^{\max},\quad\forall j\in[1:N] \tag{52}$$

$$\mathbf{0}\leq\Delta\mathbf{x}^{\max}-\Delta\mathbf{x}_j\leq M\cdot\mathbf{Z}_j,\quad\forall j\in[1:N-1] \tag{53}$$

$$\mathbf{0}\leq\Delta\mathbf{x}_{j+1}\leq M\cdot(\mathbf{1}-\mathbf{Z}_j),\quad\forall j\in[1:N-1] \tag{54}$$

$$Z_{a,j}\in\{0,1\},\quad\forall a\in\mathcal{R},j\in[1:N] \tag{55}$$

Constraints (51)-(52) provide the bounds for $\Delta\mathbf{x}_j$ . The filling order is restricted to proceed from left to right, as enforced by constraints (53)-(55). $Z_{a,j}$ is the binary variable for segment j of road a. Constraint (49) and (50) establish the linear relationship between $\mathbf{x}$ and $\Delta\mathbf{x}_j$ , and between $\mathbf{t}^{\mathrm{road}}$ and $\Delta\mathbf{x}_j$ , respectively. Constraint (11) shares the same form of constraint (10) and can be linearized using the same method.

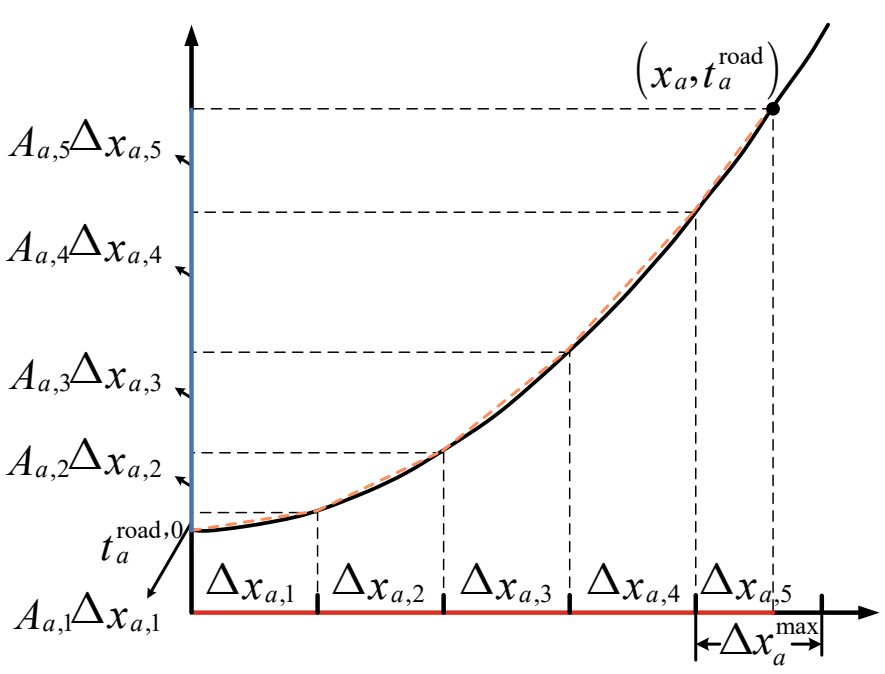

**Fig. 3.** Illustration of the five-segment piecewise linear approximation of a quartic function

### *5.2. Linearization of Constraint (17)*

We use the big-M method to replace the constraint (17) with constraints (56)-(58). Constraints (18) and (39)-(45) share the same form of constraint (17) and can be linearized using the same method.

$$\mathbf{0}\leq\mathbf{f}^{\mathrm{GV}}\leq M\cdot\mathbf{H} \tag{56}$$

$$\mathbf{0}\leq\mathbf{C}^{\mathrm{GV}}-(\boldsymbol{\Lambda}^{\mathrm{GV}})^{\mathrm{T}}\mathbf{u}^{\mathrm{GV}}\leq M\cdot(\mathbf{1}-\mathbf{H}) \tag{57}$$

$$H_c\in\{0,1\},\quad\forall c\in\rho \tag{58}$$

where $H_c$ is the binary variable for route c, with $\mathbf{H}=[H_c]_{c\in\rho}$ . $\rho=\bigcup_{rs\in\mathcal{OD}}\rho_{rs}$ is the route set. $u_{rs}^{\mathrm{GV}}$ is an auxiliary variable representing the shortest travel time for GV of O-D pair rs, with $\mathbf{u}^{\mathrm{GV}}=[u_{rs}^{\mathrm{GV}}]_{rs\in\mathcal{OD}}$ . $\mathcal{OD}$ is the set of OD-pair.

## 6. Case study

To validate and illustrate how the profit-sharing mechanism can promote the coordination of power and transportation systems, a 12-node TN and an 18-bus PDN are applied here. The topologies of TN and PDN are illustrated in Fig. 4. Twelve traffic nodes and twenty roadways make up the TN. EVCSs 1, 2, 3, 4, 5, and 6 are located at traffic node T5, T6, T7, T8, T10, and T12. The PDN consists of eighteen electric buses, seventeen lines, one substation, and three local emergency generators. EVCSs 1, 2, 3, 4, 5, and 6 are connected to electric bus B3, B12, B5, B13, B16, and 18. Generators 1, 2, and 3 are connected to bus B5, B13, and B18. The parameters of the power-transportation network are shown in Table. 1 [2].

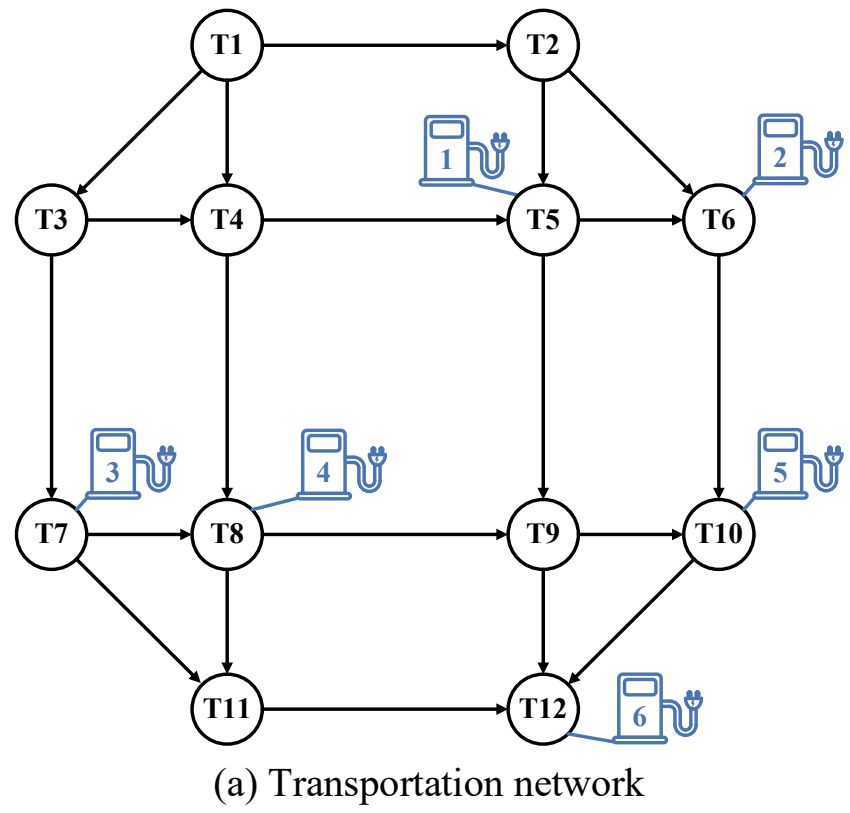

(a) Transportation network

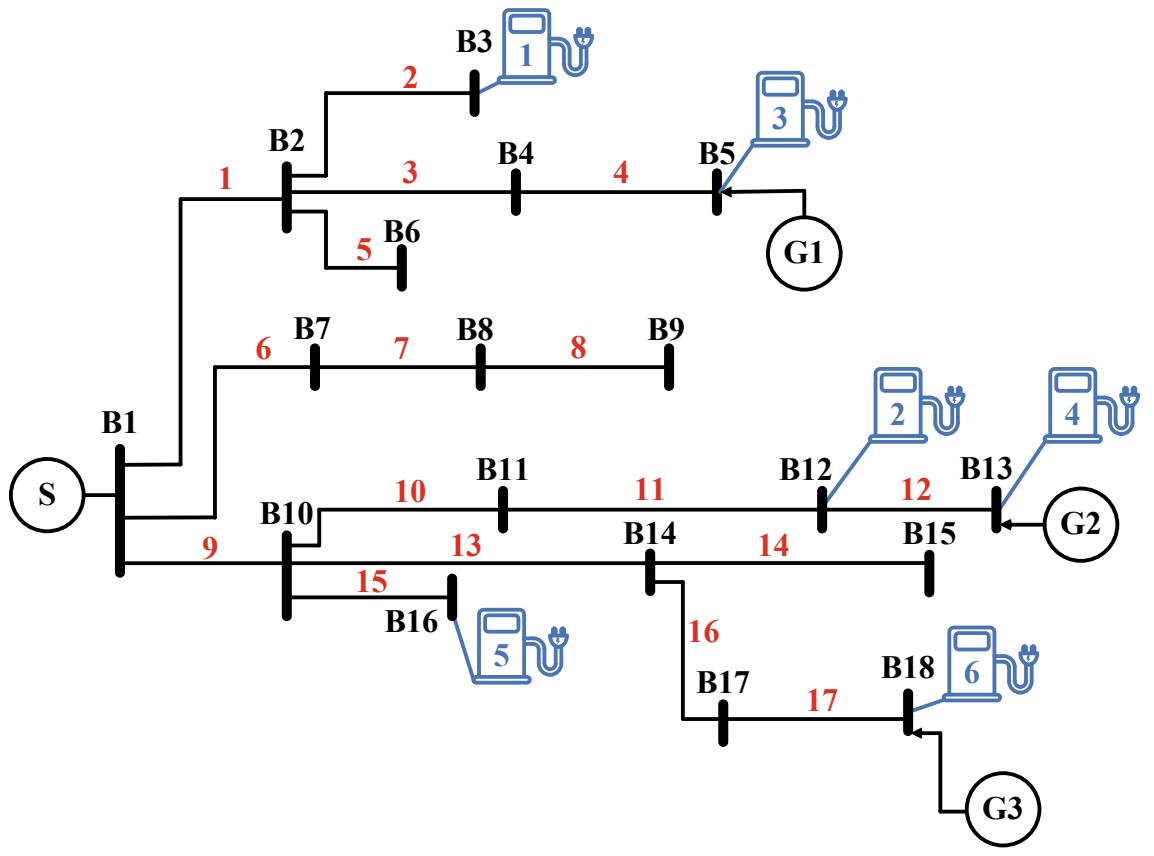

(b) Power distribution network

**Fig. 4.** Topology of TN and PDN

**Table 1.** Parameters of power-transportation network

| Power distribution network | | | Transportation network | | |
|---|---|---|---|---|---|
| parameters | value | unit | parameters | value | unit |
| $a_{1/2/3}$ | 5.2/5.4/5.8 | CNY / MW²h | $t_a^{\text{road},0}, \forall a$ | 10 | min |
| $b_{1/2/3}$ | 200/200/200 | CNY /MWh | $c_a^{\text{road}}, \forall a$ | 20 | p.u. |
| $c_{1/2/3}$ | 300/340/330 | CNY/h | $t_{1/2/3/4/5/6}^{\text{EVCS},0}$ | 30/28.5/25/ 25/27.5/25 | min |
| $\kappa$ | 400 | CNY/ MWh | $c_m^{\text{EVCS}}, \forall m$ | 12 | p.u. |
| $P^{l}_{\max,4/12/17}$ | 5.6/4.8/6.3 | MW | $J$ | 0.15 | nil |
| $X_k, \forall k$ | 0.1 | p.u. | $\omega$ | 100 | CNY /h |
| power flow base value | 100 | MVA | traffic flow base value | 100 | veh/h |
| voltage base value | 10 | kV | | | |

*6.1. Performance Evaluation of the Profit-Sharing Mechanism*

The O-D pairs and corresponding traffic demands are presented in Table 2, and the traditional electric demands are presented in Table 3. In this paper, we assume that during the scheduling period, the charging price at each EVCS is fixed at 0.6 CNY/kWh. An EV needs to select an EVCS to charge 100 kWh of electricity. GVs do not need to refuel during the journey and only require route planning.

**Table 2.** O-D pairs and traffic demands

| Vehicle Type | Gasoline vehicle | | Electric vehicle |
|---|---|---|---|
| Origin node | T1 | T5 | T1 |
| Destination node | T5 | T10 | T12 |
| Traffic Demand (veh/h) | 1600 | 1200 | 2400 |

**Table 3.** Distribution of traditional electric demands

| Bus | B6 | B8 | B9 | B10 | B15 | B16 |
|---|---|---|---|---|---|---|
| Traditional electric demand (MW) | 57.3 | 43.8 | 30.5 | 32.5 | 27.5 | 12.5 |

The simulations are conducted using MATLAB 2021a. Three cases are designed and compared:

Case I: The power and transportation systems are managed by the DNO and the TNO independently, with no coordination.

Case II: The power and transportation systems are managed by a single operator.

Case III: The power and transportation systems are managed by the DNO and the TNO independently under the profit-sharing mechanism (at the optimal sharing ratio).

**Table 4.** Cost allocations of each case

| Cases | Power distribution network (CNY) | Transportation network (CNY) | Overall cost (CNY) |
|---|---|---|---|
| I | 201621 | 162672 | 364293 |
| II | 178610 | 164764 | 343374 |
| III | Dispatch cost: 180447<br>Profit-sharing: +4235<br>Total operation cost: 184682 | Users' travel cost: 164270<br>Profit-sharing: -4235<br>Total operation cost: 160035 | 344717 |

In Case I, since there is no coordination between the TN and PDN, the TNO allocates the traffic flow and charging load without considering the operation of the PDN. The objective of the TNO is only to minimize the total travel costs of users, with a value of 162672 CNY. Under the management of the TNO, the charging load at EVCS 3, 4, and 6 is substantial. However, the capacity of line B4-B5, B12-B13, and B17-B18 is too low to transfer enough power to fully meet the charging demands at these EVCSs. Therefore, the DNO has to dispatch the local emergency generators 1, 2, and 3 to support the grid, which is uneconomic.

In Case II, the power and transportation systems are coordinated by a single operator who will only focus on minimizing the overall cost of the power-transportation system. The EVs are directed away from EVCS 3, 4, and 6 by well-designed road congestion tolls and EVCS entry fees. Finally, the charging load at these EVCSs decreases to 5.6 MW, 4.8 MW, and 6.3 MW, respectively, matching the capacities of line B4-B5, B12-B13, and B17-B18. The DNO can maintain power balance at each bus by only dispatching low marginal cost electricity from the substation. Compared with Case I, the PDN operation cost decreases by 11.41%. However, since the original optimal traffic allocation is broken to utilize the spatial flexibility of charging load, the operation cost of TN increases by 1.29%. In general, the overall cost of the power-transportation system decreases by 5.74%.

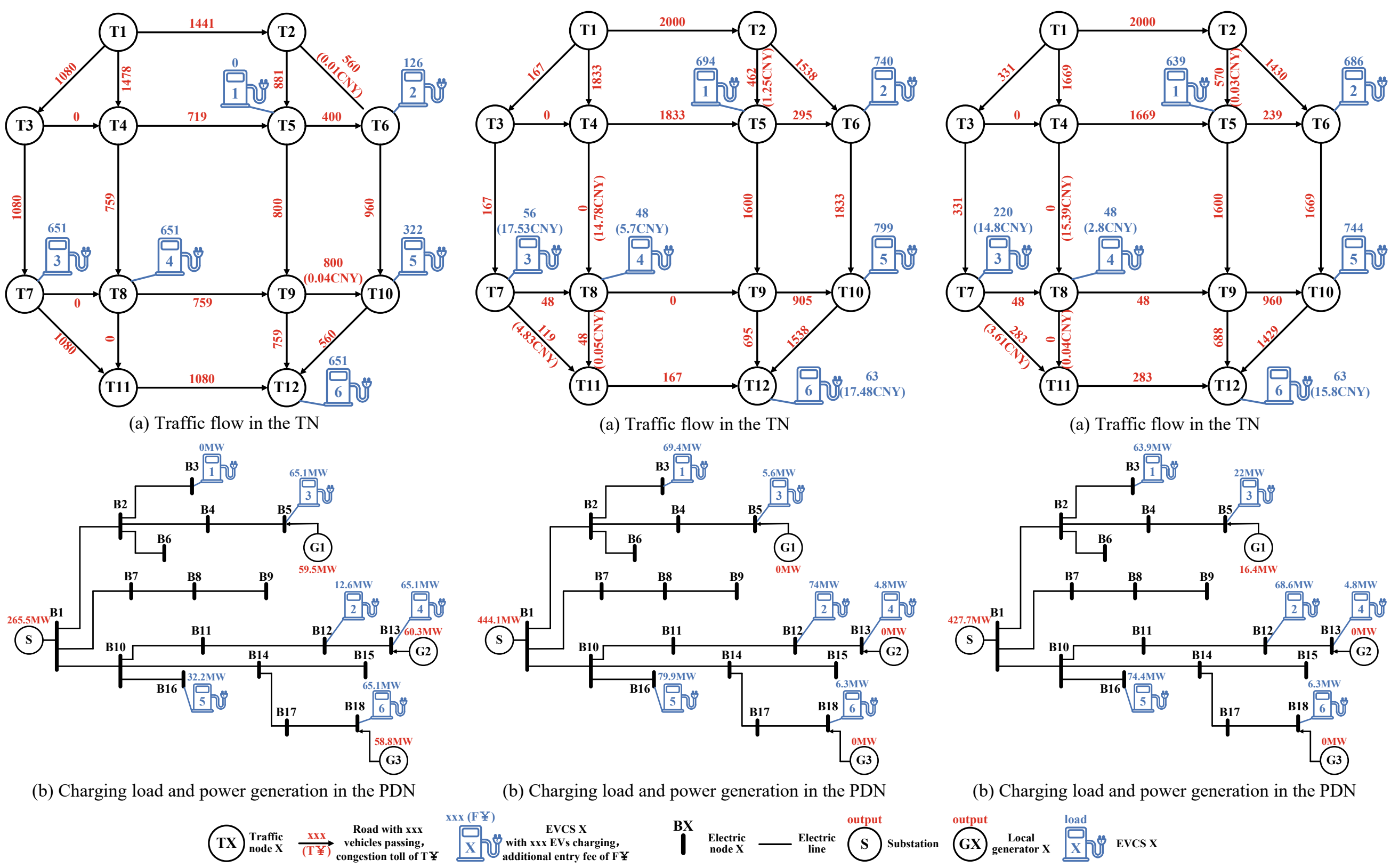

**Fig. 5.** Distribution of traffic flow, charging load, and power generation in Case I

**Fig. 6.** Distribution of traffic flow, charging load, and power generation in Case II

**Fig. 7.** Distribution of traffic flow, charging load, and power generation in Case III

Though the centralized operation enhances the efficiency of the power-transportation system, it damages the benefits of the TN. Considering the non-cooperative characteristics of the TN and the PDN in reality, the operation framework mentioned in Case II is not feasible. Therefore, we propose a profit-sharing mechanism to safeguard the benefits of both the TN and the PDN during the coordination. In Case III, the operation of power and transportation systems at the best profit-sharing ratio is researched. In this article, the optimal sharing ratio is 20%. (The specific process of determining it will be explained later in Section 6.2) Under the profit-sharing mechanism, the scheduling process of the power-transportation system is divided into two stages as pre-scheduling and re-scheduling. The situation of pre-scheduling is the same as Case I, and the benchmark operation costs of the TN and PDN are 162672 CNY ($\Gamma_0$) and 201621 CNY ($\eta_0$), respectively. At the re-scheduling stage, the TNO reallocates the traffic flow and the charging load. As a result, the charging load at EVCS 3, 4, and 6 is reduced to 22 MW, 4.8 MW, and 6.3 MW, respectively. The DNO only needs to dispatch the electricity from generator 1 and the substation. The new total travel cost of TN is 164270 CNY ($\Gamma_1$) with an increase of 1598 CNY (loss) compared with $\Gamma_0$. The new dispatch cost of PDN is 180447 CNY ($\eta_1$) with a decrease of 21174 CNY (profit) compared with $\eta_0$. The DNO needs to allocate 20% of profit to the TNO, which is 4235 CNY, to compensate for the loss of the TN. The total operation cost of TN and PDN is 160035 CNY ($\Gamma_1$ minus 4235) and 184682 CNY ($\eta_1$ plus 4235), which decreases by 1.62% and 8.4% compared with that at the pre-scheduling stage, respectively.

### *6.2. Determining the optimal sharing ratio Based on Sensitivity Analysis*

Fig. 8 demonstrates the distribution of charging load at the re-scheduling stage under different profit-sharing ratios. With the ratio increasing, much more charging load at EVCS 3, 4, 6 is transferred to EVCS 1, 2, 5. When $\alpha \geqslant 30\%$, the charging demands at EVCS 3, 4, 6 are fully matched with the capacity of line B4-B5, B12-B13, and B17-B18, respectively. Correspondingly, the power generation proportion of local generators gradually decreases. The PDN relies solely on power from the substation when $\alpha \geqslant 30\%$. (Fig. 9)

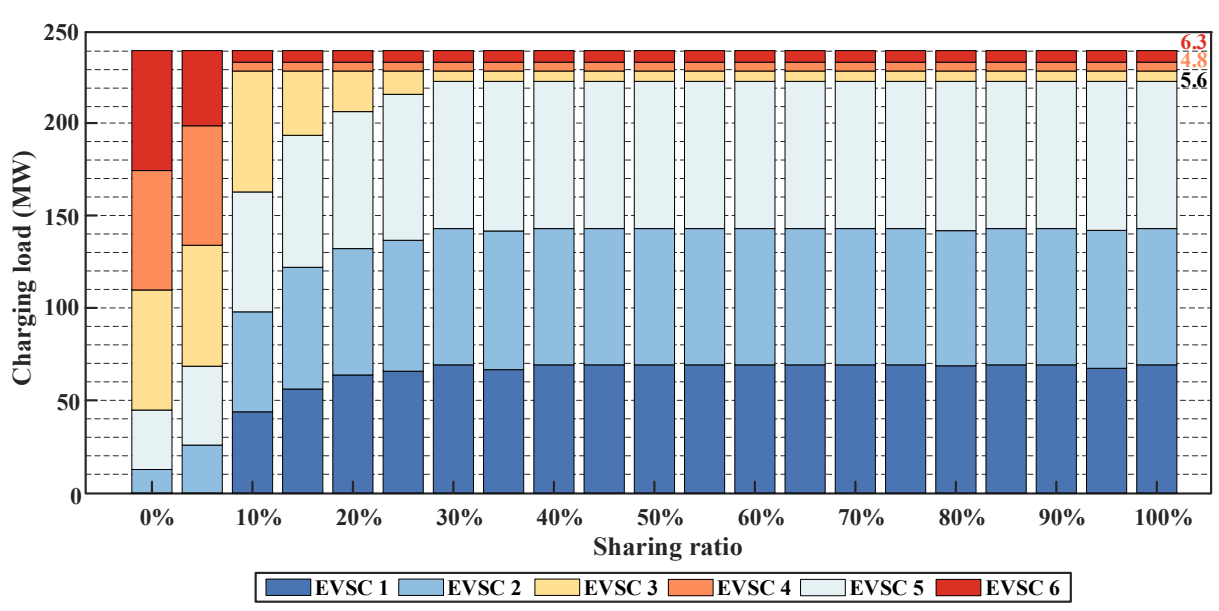

**Fig. 8.** Charging load at each EVCS under different profit-sharing ratio

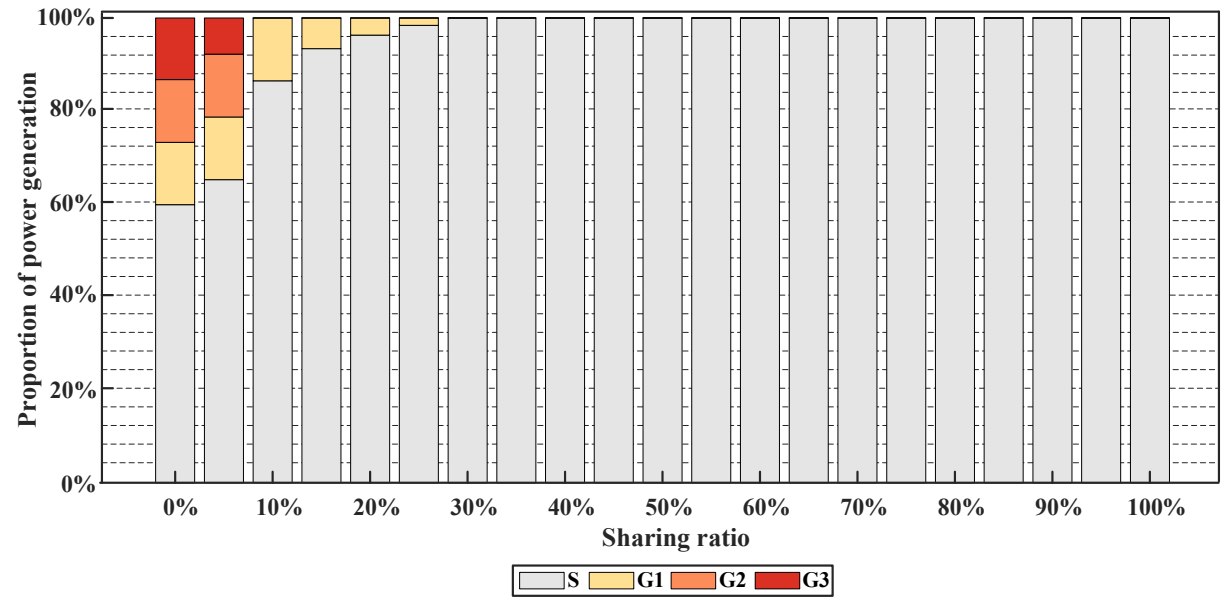

**Fig. 9.** Output proportion of substation and local generators at different profit-sharing ratios

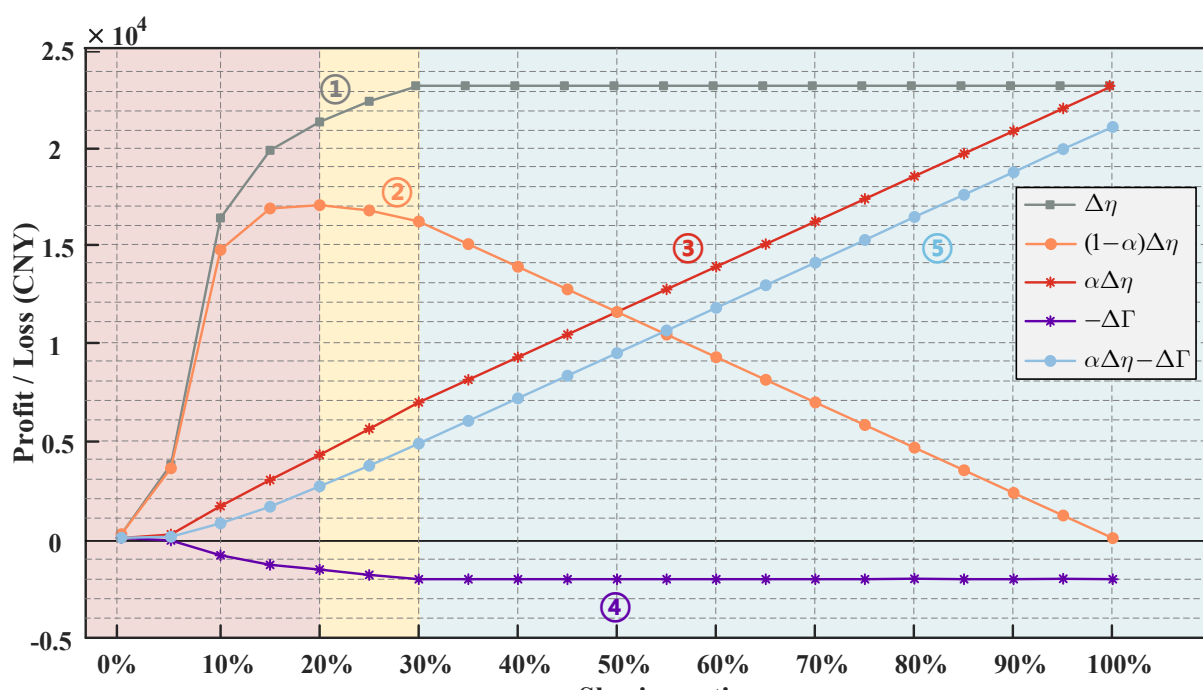

**Fig. 10.** The profit/ loss of the DNO and the TNO at different profit-sharing ratios

Due to the decrease in utilization of G1, 2, and 3 with high marginal generation costs, the PDN's dispatch cost is reduced at the re-scheduling stage. The difference between $\eta_0$ and $\eta$ is total profit $\Delta\eta$. The DNO and the TNO will receive $(1-\alpha)\Delta\eta$ and $\alpha\Delta\eta$, respectively. $(1-\alpha)\Delta\eta$ represents the net profit for the DNO. For the TNO, the reallocation leads to an increase of $\Delta\Gamma$ in the total user travel cost. We define $-\Delta\Gamma$ as the loss and the net profit of the TNO is $\alpha\Delta\eta$ plus $-\Delta\Gamma$. The trend of the profit and loss with the variation in profit-sharing ratio is shown in Fig. 10.

$$\begin{aligned}\Psi(\alpha) &= \eta_1 + \alpha\Delta\eta \\ &= (\eta_0 - \Delta\eta) + \alpha\Delta\eta \\ &= \eta_0 - (1-\alpha)\Delta\eta\end{aligned} \tag{59}$$

$$\begin{aligned}\mathrm{H}(\alpha) &= \Gamma_1 - \alpha\Delta\eta \\ &= (\Gamma_0 + \Delta\Gamma) - \alpha\Delta\eta \\ &= \Gamma_0 - (-\Delta\Gamma + \alpha\Delta\eta)\end{aligned} \tag{60}$$

At the re-scheduling stage, the total cost of the DNO and the TNO mentioned in (27) and (29) can be transformed by equations (59)-(60). (Benchmark cost minus the net profit) The trend of them with the variation in profit-sharing ratio is shown in Fig. 11. According to constraint (48), the best profit-sharing ratio is determined as 20%.

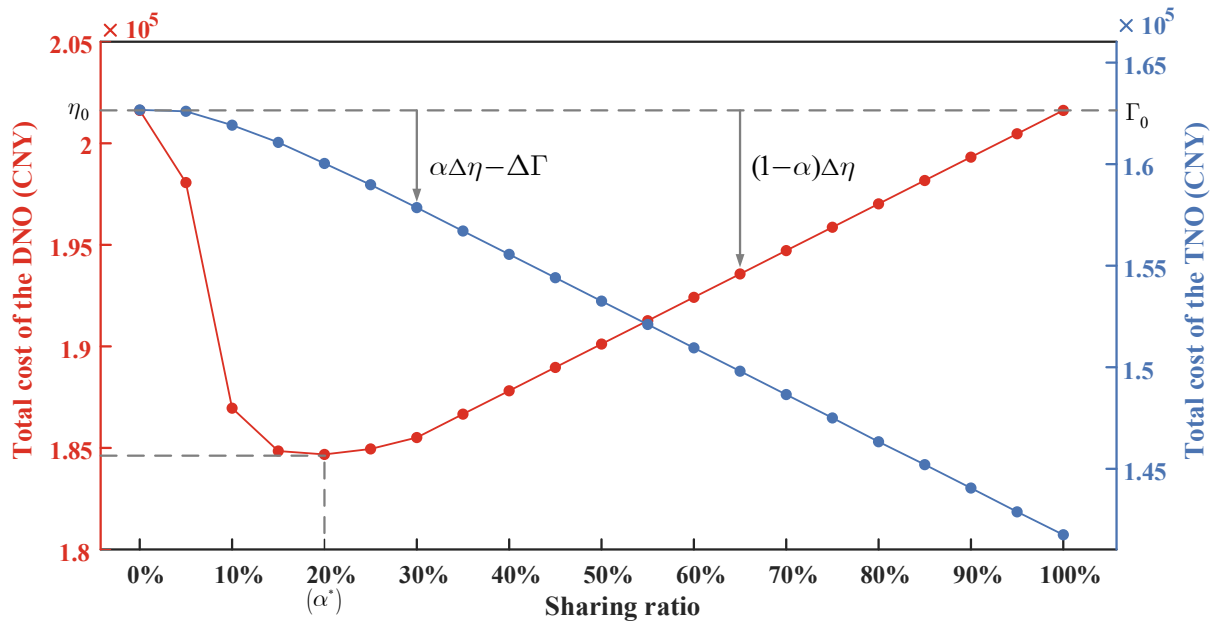

**Fig. 11.** The total scheduling cost of the DNO and the TNO at different profit-sharing ratios

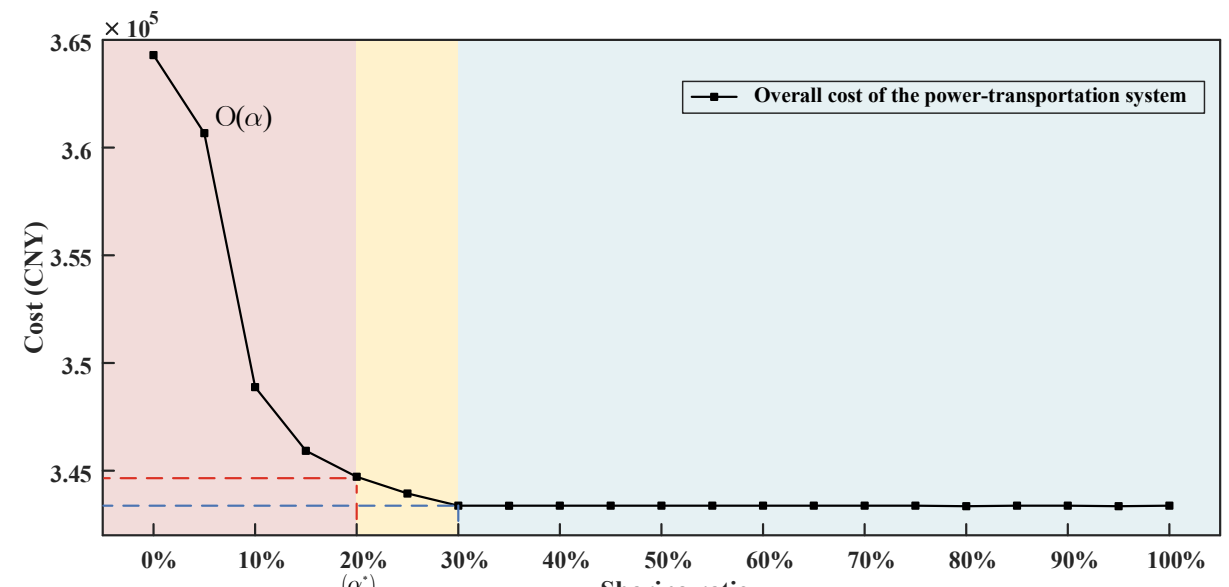

**Fig. 12.** Overall cost of the power-transportation system under different profit-sharing ratios

Region I & Region II: With the profit-sharing ratio increasing, the regulatory potential on the traffic side is gradually stimulated. On one hand, the reallocation of charging load alleviates the mismatch between the electricity supply and demand in the PDN, which brings profit ($\Delta\eta$). On the other hand, there is an increase in traffic users' travel cost ($\Delta\Gamma$). Since the marginal increase of $\Delta\eta$ (curve ①) is faster than that of $\Delta\Gamma$ (curve ④), the discriminant (61) is fulfilled. Therefore, the overall cost of the power-transportation system monotonically decreases with $\alpha$.

$$\frac{\mathrm{dO}(\alpha)}{\mathrm{d}\alpha} = \frac{\mathrm{d}(\mathrm{H}(\alpha)+\Psi(\alpha))}{\mathrm{d}\alpha} = \frac{\mathrm{d}\Delta\Gamma}{\mathrm{d}\alpha} - \frac{\mathrm{d}\Delta\eta}{\mathrm{d}\alpha} < 0 \tag{61}$$

Region III: The regulatory potential on the traffic side is fully stimulated. The state of the power-transportation system is the same as that of Case II (centralized operation), including the distribution of power flow, traffic flow, and charging load. The overall cost of the power-transportation system remains at the lowest level. If the target is just to maximize the social welfare, we can choose any value in [0.3,1] as the profit-sharing ratio.

In this article, we focus on minimizing the total operation cost of the PDN under the profit-sharing mechanism. Therefore, regions II and III are not considered, though the overall cost of the power-transportation system can be further decreased. The reason why the sharing ratios promoting the minimal operation cost of PDN and the power-transportation system are different is the non-cooperative relationship between the TNO and the DNO.

## 7. Conclusion

This paper proposes a profit-sharing mechanism to promote the coordination between the TN and the PDN. Specifically, the profit-sharing can drive the TNO to utilize the spatial flexibility of EV charging load to improve the security and economy of the PDN. The proposed bilevel model can determine the profit-sharing ratio that is most beneficial to minimize the total operation cost of the PDN. The simulations are based on a coupled power-transportation network, in which the TN contains twenty roadways and six EVCSs, and the PDN consists of seventeen lines, one substation, and three local emergency generators. The results demonstrate that the profit-sharing mechanism can promote the efficient operation of the non-cooperative power-transportation system. In detail, at the optimal sharing ratio (In this article, it is 20%), the TNO reallocates the charging load to help the DNO reduce the utilization of the local generators by 90.82%. The total operation cost of the PDN and TN can be reduced by 8.4% and 1.62%, respectively.

As long as the two systems A and B satisfy the following properties, the profit-sharing mechanism proposed in this paper can be used to coordinate their operations: (a) System A and system B are non-cooperative during the scheduling process. (b) Leveraging the regulatory flexibility of system A can benefit the operation of system B. (c) The utilization of the flexibility simultaneously harms the interests of system A. Here, system A represents a flexible load system whose operating schedule can be temporally shifted or re-dispatched without violating critical constraints, such as steel industrial loads (e.g., modifying steelmaking schedules, delaying batch processing in electric arc furnaces) or data center computing loads (e.g., postponing non-real-time batch jobs, transferring computing tasks across geographically distributed servers). System B, on the other hand, represents the power system operator that can benefit from the demand-side flexibility provided by system A for purposes such as peak shaving, congestion relief, or renewable energy accommodation. In future work, the method that the TNO allocates the benefit shared by the DNO to each EV user will be investigated to extend the current mechanism.

## References

[1] International Energy Agency (IEA), Global EV Outlook 2024, IEA, Paris, 2024. https://www.iea.org/reports/global-ev-outlook-2024 (accessed May 20, 2026).

[2] W. Gan, J. Wen, M. Yan, Y. Zhou and W. Yao, "Enhancing resilience with electric vehicles charging redispatching and vehicle-to-grid in traffic-electric networks," *IEEE Trans. Industry Applications*, vol. 60, no. 1, pp. 953-965, Jan.-Feb. 2024.

[3] M. Singh, I. Kar and P. Kumar, "Influence of EV on grid power quality and optimizing the charging schedule to mitigate voltage imbalance and reduce power loss," Proceedings of 14th International Power Electronics and Motion Control Conference EPE-PEMC 2010, Ohrid, Macedonia, 2010.

[4] GA Putrus, Pasist Suwanapingkarl, David Johnston, EC Bentley, and Mahinsasa Narayana, "Impact of electric vehicles on power distribution networks," 2009 IEEE Vehicle Power and Propulsion Conference, pages 827–831. IEEE, 2009.

[5] L. Pieltain Fernández, T. Gomez San Roman, R. Cossent, C. Mateo Domingo and P. Frías, "Assessment of the Impact of Plug-in Electric Vehicles on Distribution Networks," *IEEE Trans. Power Systems*, vol. 26, no. 1, pp. 206-213, Feb. 2011.

[6] S. Acha, T. C. Green and N. Shah, "Effects of optimised plug-in hybrid vehicle charging strategies on electric distribution network losses," IEEE PES T&D 2010, New Orleans, LA, USA, 2010, pp. 1-6.

[7] W. Gan et al., "Coordinated planning of transportation and electric power networks with the proliferation of electric vehicles," *IEEE Trans. Smart Grid*, vol. 11, no. 5, pp. 4005-4016, Sept. 2020.

[8] W. Gan et al., "A tri-level planning approach to resilient expansion and hardening of coupled power distribution and transportation systems," *IEEE Trans. Power Systems*, vol. 37, no. 2, pp. 1495-1507, March 2022.

[9] B. Mukherjee and F. Sossan, "Optimized planning of chargers for electric vehicles in distribution grids including PV self - consumption and cooperative vehicle owners," *Energy Conversion and Economics*, vol. 4, no. 1, pp. 36-46, 2023.

[10] Y. Sheffi, *Urban Transportation Networks: Equilibrium Analysis With Mathematical Programming Methods*. Englewood Cliffs, NJ, USA: Prentice-Hall, 1985.

[11] W. Wei, S. Mei, L. Wu, M. Shahidehpour and Y. Fang, "Optimal traffic-power flow in urban electrified transportation networks," *IEEE Trans. Smart Grid*, vol. 8, no. 1, pp. 84-95, Jan. 2017.

[12] W. Gan et al., "Multi-network coordinated hydrogen supply infrastructure planning for the integration of hydrogen vehicles and renewable energy," *IEEE Trans. Industry Applications*, vol. 58, no. 2, pp. 2875-2886, March-April 2022.

[13] W. Wei, L. Wu, J. Wang and S. Mei, "Network equilibrium of coupled transportation and power distribution systems," *IEEE Trans. Smart Grid*, vol. 9, no. 6, pp. 6764-6779, Nov. 2018.

[14] M. Yan, M. Shahidehpour, A. Paaso, L. Zhang, A. Alabdulwahab and A. Abusorrah, "Distribution system resilience in ice storms by optimal routing of mobile devices on congested roads," *IEEE Trans. Smart Grid*, vol. 12, no. 2, pp. 1314-1328, March 2021.

[15] J. Wen, W. Gan, C. -C. Chu, L. Jiang and J. Luo, "Robust resilience enhancement by EV charging infrastructure planning in coupled power distribution and transportation systems," *IEEE Trans. Smart Grid*, vol. 16, no. 1, pp. 491-504, Jan. 2025.

[16] Y. Cui, Z. Hu and X. Duan, "Optimal pricing of public electric vehicle charging stations considering operations of coupled transportation and power systems," *IEEE Trans. Smart Grid*, vol. 12, no. 4, pp. 3278-3288, July 2021.

[17] H. Xiong *et al*., "Dynamic programming-based non-anticipative approach for multi-stage coordinated optimal dispatch of electricity and transportation nexus," *IEEE Trans. Smart Grid*, doi: 10.1109/TSG.2026.3687884.

[18] S. Lv, S. Chen and Z. Wei, "Coordinating urban power-transportation networks: a subsidy-based Nash–Stackelberg–Nash game model," *IEEE Trans. Industrial Informatics*, vol. 19, no. 2, pp. 1778-1790, Feb. 2023.

[19] K. Li, C. Shao, M. Shahidehpour and X. Wang, "A capacity-based regulation method for coordinating electric vehicle charging flows in coupled distribution and transportation networks," *IEEE Trans. Smart Grid*, vol. 15, no. 3, pp. 3066-3079, May 2024.

[20] T. Qian, M. Fang, Q. Hu, C. Shao and J. Zheng, "V2Sim: an open-source microscopic V2G simulation platform in urban power and transportation network," *IEEE Trans. Smart Grid*, vol. 16, no. 4, pp. 3167-3178, July 2025.

[21] Fuzhang Wu, Jun Yang, et al, "Coordinated fault risk prevention in coupled distribution and transportation networks considering flexible travel demands," *Protection and Control of Modern Power Systems*, 2025, V10(04):16-27.

[22] Padhmanabhaiyappan Sivalingam, Madhusudanan Gurusamy. Momentum Search Algorithm for Analysis of Fuel Cell Vehicle-to-Grid System with Large-Scale Buildings," *Protection and Control of Modern Power Systems*, 2024, V9(2):147-160.

[23] T. Qian, C. Shao, X. Li, X. Wang and M. Shahidehpour, "Enhanced coordinated operations of electric power and transportation networks via EV charging services," *IEEE Tran. Smart Grid*, vol. 11, no. 4, pp. 3019-3030, July 2020.

[24] S. Lv, Z. Wei, G. Sun, S. Chen and H. Zang, "Optimal Power and Semi-Dynamic Traffic Flow in Urban Electrified Transportation Networks," *IEEE Trans. Smart Grid*, vol. 11, no. 3, pp. 1854-1865, May 2020.

[25] Y. Liu *et al*., "Coordinated dispatch of power and transportation systems considering hydrogen storage based on heterogeneous decomposition," *IEEE Trans. Transportation Electrification*, vol. 10, no. 3, pp. 7526-7539, Sept. 2024.

[26] Vohra RV. *Mechanism Design: A Linear Programming Approach*. Econometric Society Monographs. Cambridge University Press; 2011:1-6.

[27] M. Alizadeh, H. -T. Wai, M. Chowdhury, A. Goldsmith, A. Scaglione and T. Javidi, "Optimal pricing to manage electric vehicles in coupled power and transportation networks," *IEEE Trans. Control of Network Systems*, vol. 4, no. 4, pp. 863-875, Dec. 2017.

[28] T. A. Manual, *Bureau of Public Roads*, U.S. Dept. Commission, Washington, DC, USA, 1964.

[29] K. Davidson, "A flow travel time relationship for use in transportation planning," in Proc. 3rd Aust. Road Res. Board (ARRB) Conf., vol. 3, 1966, pp. 183–194.

[30] M. Yan, N. Zhang, X. Ai, M. Shahidehpour, C. Kang and J. Wen, "Robust two-stage regional-district scheduling of multi-carrier energy systems with a large penetration of wind power," *IEEE Trans. Sustainable Energy*, vol. 10, no. 3, pp. 1227-1239, July 2019.

[31] F. Li and R. Bo, "DCOPF-based LMP simulation: algorithm, comparison with ACOPF, and sensitivity," *IEEE Trans. Power Systems*, vol. 22, no. 4, pp. 1475-1485, Nov. 2007.

[32] Y. Zhao, Y. Cao, Y. Li, B. Zeng, M. Shahidehpour and Y. Cai, "Risk-based contingency screening method considering cyber-attacks on substations," *IEEE Trans. Smart Grid*, vol. 13, no. 6, pp. 4973-4976, Nov. 2022.

## Acknowledgment

This work was supported by the National Natural Science Foundation of China under Grant 52307100 and Grant W2512030.